\pdfoutput=1
\documentclass[11pt]{article}

\usepackage{PRIMEarxiv}

\usepackage{times}
\usepackage{latexsym}
\usepackage{url}
\usepackage{placeins}
\usepackage{float}
\usepackage{placeins}
\usepackage{amsmath}

\usepackage[T1]{fontenc}
\usepackage[utf8]{inputenc}

\usepackage{microtype}

\usepackage{listings}
\usepackage{xcolor}

\lstdefinestyle{python}{
  language=Python,
  basicstyle=\ttfamily\small,
  columns=fullflexible,
  keepspaces=true,
  showstringspaces=false,
  breaklines=true,
  breakatwhitespace=true,
  frame=single,
  numbers=left,
  numberstyle=\tiny,
  xleftmargin=2em,
  framexleftmargin=1.5em,
  tabsize=4
}
\usepackage{inconsolata}
\usepackage{textcase}  % in your preamble

\usepackage{graphicx}

\usepackage{xcolor}
\usepackage{courier}  % For \texttt{}
\usepackage{textcase} % For \MakeTextLowercase

\definecolor{darkgreen}{HTML}{057201}

\title{\NoCaseChange{\textcolor{darkgreen}{irpapers}}: A Visual Document Benchmark for Scientific Retrieval and Question Answering}

\author{
    Connor Shorten \\
    Weaviate \\
    \And
    Augustas Skaburskas \\
    Weaviate \\
    \And
    Daniel M. Jones \\
    Weaviate \\
    \And
    Charles Pierse \\
    Weaviate \\
    \AND
    Roberto Esposito \\
    Weaviate \\
    \And
    John Trengrove \\
    Weaviate \\
    \And
    Etienne Dilocker \\
    Weaviate \\
    \And
    Bob van Luijt \\
    Weaviate \\
}

\begin{document}

\maketitle

\begin{abstract}
AI systems have achieved remarkable success in processing text and relational data, however, visual document processing remains relatively underexplored. Whereas traditional systems require OCR transcriptions to convert these visual documents into text and metadata, recent advances in multimodal foundation models offer an alternative path: retrieval and generation directly from document images. This raises a timely and important question: \textit{How do image-based systems compare to established text-based methods?} To answer this question, we present \textcolor{darkgreen}{IRPAPERS}, a benchmark totaling 3,230 pages sourced from 166 scientific papers, with both an image and OCR transcription for each page. We present a curation of 180 needle-in-the-haystack questions for evaluating retrieval and question answering systems with this corpus. We begin by comparing image- and text-based retrieval with open-source models, as well as multimodal hybrid search. For image retrieval, we evaluate the ColModernVBERT multi-vector embedding model. For text retrieval, we evaluate Arctic 2.0 dense single-vector embeddings, BM25, and their combination in hybrid text search. Text-based methods achieved 46\% Recall@1, 78\% Recall@5, and 91\% Recall@20, while image-based retrieval achieved 43\% Recall@1, 78\% Recall@5, and 93\% Recall@20. These retrieval systems exhibit complementary failures, each succeeding on queries where the other fails, enabling multimodal fusion to exceed either modality alone. Multimodal hybrid search achieved the highest performance with 49\% Recall@1, 81\% Recall@5, and 95\% Recall@20. We additionally evaluate the efficiency-performance tradeoff of MUVERA encoding with varying levels of $ef$, as well as the performance of the ColPali and ColQwen2 multi-vector image embeddings models. To contextualize open-source performance, we further evaluate leading closed-source models. Cohere Embed v4 page image embeddings reached 58\% Recall@1, 87\% Recall@5, and 97\% Recall@20, outperforming Voyage 3 Large text transcription embeddings at 52\% Recall@1, 86\% Recall@5, and 95\% Recall@20, as well as all tested open-source models. For question answering, we evaluate RAG using an LLM-as-Judge to assess binary semantic equivalence between the ground truth and the answer from the tested RAG system. RAG systems using text inputs achieved a ground truth alignment score of 0.82, compared to 0.71 for image inputs. Notably, both modalities benefit substantially from increased retrieval depth for question answering. Retrieving five documents outperforms even oracle single-document retrieval, suggesting that related pages provide valuable supporting context for answer synthesis. We conclude by exploring the comparative limitations of text and image unimodal representations: \textit{Are there questions that require the image representation to answer?}, and inversely, \textit{Are there questions that require a text representation to answer?}. These results illuminate the current state of visual document search and question answering technologies. We have open-sourced our \textcolor{darkgreen}{IRPAPERS} dataset on HuggingFace at \url{huggingface.co/weaviate/IRPAPERS} and GitHub at \url{github.com/weaviate/IRPAPERS}. Our experimental code is also available on GitHub at \url{github.com/weaviate/query-agent-benchmarking}.
\end{abstract}

\section{Introduction}

Visual document processing has seen rapid and sustained progress in recent years. Advances in multimodal representation learning now allow AI systems to retrieve from and read directly over document images. Rather than relying on OCR-derived text and metadata representations, these systems process native image representations. This progress is driven by two complementary trends. First, multimodal foundation models now embed document images and text queries into a unified representation space, from single-vector to late-interaction architectures, and can read document images directly for downstream reasoning. Second, advances in retrieval systems have enabled efficient indexing, approximate encoding of multi-vector representations, and hybrid fusion strategies. Together, these developments motivate us to revisit a fundamental question: \textit{How do image-based systems compare to established text-based systems for scientific retrieval and question answering?}

\begin{figure}
    \centering
    \includegraphics[width=1.0\linewidth]{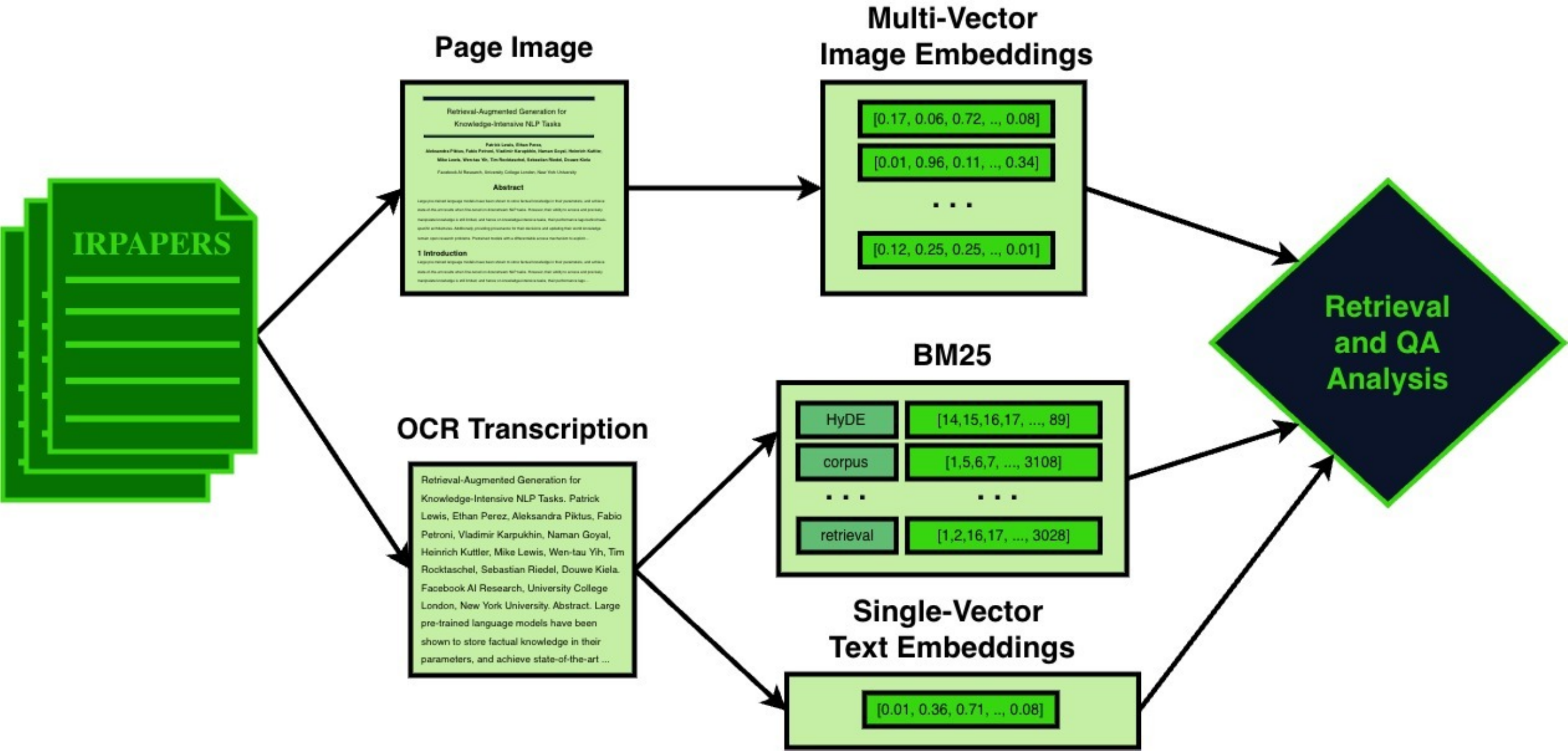}
    \caption{An overview of our experimental methodology. We compare retrieval systems operating on embedded page images with ColModernVBERT and MUVERA encoding to embedded text transcriptions that leverage GPT-4.1 OCR and a hybrid search retrieval strategy combining Arctic 2.0 dense text embeddings with BM25. We find the highest performance with multimodal hybrid search, combining normalized scores from all three retrieval systems pictured.}
    \label{fig:placeholder}
\end{figure}

Scientific literature is the primary substrate through which scientific knowledge accumulates. Every new result seeks to be contextualized against prior work, yet researchers today face an unprecedented volume of publications that exceeds individual capacity to search, read, and aggregate. In computer science alone, approximately 300 new papers appear on arXiv daily, and venues such as NeurIPS now publish over 5,000 papers annually. Effective retrieval and question answering over this literature will increasingly shape how novelty is assessed, how ideas propagate, and ultimately how science advances. As AI systems increasingly assist or autonomously participate in scientific workflows, their ability to retrieve and reason over the scientific record becomes a foundational capability. Scientific papers are also inherently visual documents. Beyond prose, they encode meaning through layout, figures, equations, and diagrams. Whether these visual cues are faithfully preserved when documents are transcribed to text remains an open question, making scientific literature a natural testbed for evaluating whether image-based representations preserve signals lost through text transcription.

We introduce \textcolor{darkgreen}{IRPAPERS}, a visual document benchmark comprising 3,230 pages from 166 scientific papers and 180 questions designed to evaluate visual document retrieval and question answering. The papers are sourced from the citations of a comprehensive survey of large language models for information retrieval \cite{zhu}. Survey papers provide a curated and structured view of prior work, grouping related methods and assumptions into coherent lines of comparison. By drawing from a survey’s citations, \textcolor{darkgreen}{IRPAPERS} yields a semantically dense corpus in which documents share vocabulary, techniques, and experimental settings. For example, a query such as: \textit{In HyDE, what instruction-following model is used to generate hypothetical documents, and which contrastive encoder is applied for non-English retrieval?} requires discriminating among multiple dense retrieval papers that discuss similar techniques but differ in specific architectural and training choices. This tests whether retrieval systems can discriminate among papers that discuss similar concepts and whether question answering systems can parse through multiple related evidence. Unlike broad scientific corpora spanning all disciplines, \textcolor{darkgreen}{IRPAPERS} enables rapid iteration and reproducibility without large infrastructure overhead. Further, the number of pages used in \textcolor{darkgreen}{IRPAPERS} is on par with the subsets used in ViDoRe \cite{vidorev2, vidorev3}, one of the most popular benchmarks for visual document retrieval at the time of this publication.

Our benchmark enables direct comparison between image- and text-based approaches for both retrieval and question answering over scientific literature. For retrieval, open-source image embeddings achieve performance comparable to hybrid text search at the top rank, 43\% vs. 46\% Recall@1, while matching or exceeding text-based methods at deeper recall. Interestingly, text and image representations exhibit complementary failure modes. At Recall@1, 22 queries succeed with text, but fail with images, while 18 succeed with images, but fail with text. Multimodal hybrid search exploits this complementarity by fusing scores from both modalities, achieving 49\% Recall@1. Scaling image-based retrievers further improves top-ranked accuracy: larger multi-vector models such as ColPali and ColQwen2 reach up to 49\% Recall@1, while yielding similar Recall@20 to smaller models, indicating diminishing returns at deeper ranking depths. We additionally evaluate MUVERA encoding, and demonstrate a clear performance-efficiency tradeoff with hyperparameter values of ef. With ef set to 1024, we reach 41\% Recall@1, a 2-point absolute drop from ColModernVBERT retrieval without MUVERA encoding. To contextualize open-source performance, we evaluate leading closed-source models. Cohere Embed v4 multimodal embeddings achieve 58\% Recall@1 on page images, outperforming Voyage 3 Large text embeddings at 52\% Recall@1 and establishing a 9-point absolute Recall@1 gap over the best open-source approach.

For question answering, we begin by grounding the benchmark with No Retrieval and Hard Negative baselines to validate question difficulty and assess false negative labeling in the retrieval evaluation. Without retrieval, models achieve an alignment score of 0.16, indicating that questions cannot be reliably answered from parametric knowledge alone. Providing hard negative context further demonstrates the challenge of the task. Hard Negative image context reduces performance to 0.12, and Hard Negative text context reaches only 0.39. Together, these results validate that the benchmark requires precise retrieval rather than general knowledge or topical similarity. TextRAG consistently outperforms ImageRAG at both k = 1 and k = 5, achieving 0.62 vs. 0.40 at k = 1 and 0.82 vs. 0.71 at k = 5. Oracle retrieval provides a reference point for single-document evidence, achieving 0.74 alignment for text-based question answering and 0.68 for image-based question answering. Notably, performance at k=5 exceeds oracle single-document retrieval for both text and image inputs, indicating that scientific question answering often requires synthesizing complementary evidence across multiple related pages rather than relying on a single definitive source.

Unlike plain text, visual documents encode meaning through deliberate spatial arrangement. A central question raised by \textcolor{darkgreen}{IRPAPERS} is how representational choice shapes what information can be recovered in scientific information retrieval. More generally, prior work on system design shows that representations determine which classes of information from symbolic, relational, or structural, are preserved or lost downstream \cite{ontology}. We therefore conclude by exploring an explicit comparison between image- and text-based AI systems: \textit{Do image representations contain answers to questions that would be impossible to answer with text transcriptions?} We identify four categories of visual elements contained in \textcolor{darkgreen}{IRPAPERS}: Architectural Diagrams, Tables, Charts, and Abstract Concepts, and construct two sets of questions intended to isolate cases where visual grounding is required. We conclude by discussing agentic search systems that decompose queries into semantic retrieval and structured constraints as a promising direction for future work, addressing the inverse question: \textit{Do text representations contain answers to questions that would be impossible to answer with image representations?}

Our contributions are as follows:

\begin{itemize}
    \item We release \textcolor{darkgreen}{IRPAPERS}, a benchmark comprising 166 information retrieval papers (3,230 pages) with 180 curated queries targeting precise methodological details.
    \item We present a systematic comparison of multi-vector image retrieval against hybrid text search for scientific documents with open-source models. We demonstrate that multimodal hybrid search combining open-source text and image models outperforms unimodal baselines.
    \item We evaluate MUVERA encoding for multi-vector image embeddings, quantifying its tradeoff between efficiency and retrieval quality with varying levels of $ef$.
    \item We evaluate leading closed-source embedding models, Cohere Embed v4.0 and Voyage 3 Large, quantifying the current performance gap between open-source and closed-source approaches on visual document retrieval.
    \item We compare retrieval-augmented generation across modalities, finding that both benefit substantially from increased retrieval depth, with five retrieved documents outperforming oracle single-document context.
\end{itemize}

\section{\textcolor{darkgreen}{IRPAPERS} Dataset}

\begin{table}[h]
\centering
\label{tab:trainable_rewriter_qa}
\setlength{\tabcolsep}{12pt}
\renewcommand{\arraystretch}{1.5}
\begin{tabular}{|p{6cm}|p{8cm}|}
\hline
\textbf{Question} & \textbf{Answer} \\
\hline
In HyDE, what specific instruction-following models and contrastive encoders were used for English versus non-English retrieval tasks? &
HyDE uses InstructGPT for all tasks, Contriever for English retrieval tasks, and mContriever for non-English tasks. \\
\hline
How does HyDE perform on the Arguana dataset compared to BM25 and ANCE in terms of nDCG@10? & HyDE achieves 46.6 nDCG@10 on Arguana, outperforming both BM25 (39.7) and ANCE (41.5). \\
\hline
In the paper "Generative and Pseudo-Relevant Feedback for Sparse, Dense and Learned Sparse Retrieval", what LLM-based approach generates text independent of first-pass retrieval effectiveness, and how many diverse types of generated content does it produce? &  Generative-relevance feedback (GRF) uses GPT-3 to generate ten diverse types of text (including chain-of-thought reasoning, facts, news articles, etc.) that act as "generated documents" for term-based expansion models, independent of first-pass retrieval. \\
\hline
\end{tabular}
\vspace{1em}
\caption{Question, Answer samples from \textcolor{darkgreen}{IRPAPERS}.}
\end{table}

We introduce \textcolor{darkgreen}{IRPAPERS}, a benchmark dataset of 166 scientific papers focused on Information Retrieval (IR), spanning 3,230 pages. Unlike large-scale scientific corpora such as S2ORC \cite{s2orc}, which prioritize breadth across all disciplines, \textcolor{darkgreen}{IRPAPERS} is constrained to a single research community. This reflects realistic workflows with scientific literature, researchers rarely search across all of science, but instead navigate tightly interconnected subfields. Our collection is sourced from papers cited in ``Large Language Models for Information Retrieval: A Survey'' by Zhu et al. \cite{zhu}, an influential work with 606 citations since its original publication in 2023. For example, \textcolor{darkgreen}{IRPAPERS} contains works on reranking, such as Rank1 and RankT5 \cite{rank1, rankt5}, and query expansion, such as HyDE and Doc2Query \cite{hyde, doc2query}, among other IR subfields. This semantic density creates a challenging discrimination task, retrieval systems cannot rely on surface-level topic differences and must instead identify fine-grained methodological distinctions. \textcolor{darkgreen}{IRPAPERS} could provide AI assistants with grounded access to foundational IR research. The evaluation we present establishes how well current systems surface precise methodological details from this corpus, building trust for practical deployment and a foundation for continued improvement.

We construct questions from 19 papers focused on Query Writing, totaling 180 questions across non-reference pages. These questions are generated following the needle-in-the-haystack benchmark philosophy \cite{lostinthemiddle}: \textit{Can a retrieval system find the source document used to create the test question?} For each of the papers in our document corpus, we feed it to the Claude chat application using Claude Sonnet 4.5 \cite{claude45sonnet} with the prompt shown in Appendix A.2. We construct one question for each page per paper in our document corpus, skipping pages that contain only references. Table 1 illustrates representative question–answer pairs from \textcolor{darkgreen}{IRPAPERS}. For example, a query such as: \textit{In HyDE, what specific instruction-following models and contrastive encoders were used for English versus non-English retrieval tasks?}, must discriminate among several papers that reference the HyDE methodology. We validate query discriminativeness in Section 4.2 by demonstrating that providing RAG systems with the top-ranked non-gold document, commonly referred to as a hard negative in training algorithms, yields substantially degraded performance. While this does not eliminate all possible false negatives, the substantial degradation under hard negative retrieval suggests that most questions require the specific source page to answer.

Our benchmark compares two practical representations of PDFs: page images and transcribed text. For the image representation, we split each PDF into its constituent page images and store them as base64-encoded strings. At an average of 1.3 MB per page, the resulting storage is 4.2 GB for 3,230 pages. This preprocessing step is simple and deterministic. On an Apple M1 Pro, base64 encoding takes 130 ms per page, corresponding to 420 seconds of total CPU work to process the entire corpus of 3,230 pages. With bounded parallelism and streaming writes to avoid buffering the full corpus in memory, the theoretical lower bound with 8 workers is 52.5 seconds wall-clock, with realized wall-clock depending on I/O and runtime overheads.

For the text representation, we transcribe each page image using OpenAI’s GPT-4.1 Multimodal Foundation Model API \cite{GPT4} with the prompt shown in Appendix A.1. With our experimental setup, we observe an inference latency of 25 seconds per page on average, utilizing 1,081 input tokens, 1,125 output tokens per page, and 2,206 tokens per page in total. With UTF-8 encoding, the 1,125 output tokens required for storing the text representation per page correspond to about 4.5 KB. Compared to 1.3 MB per page for base64-encoded images, text transcriptions are about 290x cheaper to store than page images. Although the resulting text is cheaper to store than page images, transcription introduces computational and operational overhead. At an entry-level API tier capped at 30,000 tokens per minute, this corresponds to about 13 pages per minute and 4 hours for the full corpus. At the time of this writing, GPT-4.1 is priced at \$3.00 per million input tokens and \$12.00 per million output tokens. Under these rates and the observed token counts, transcription costs \$0.017 per page and \$54.08 for the full \textcolor{darkgreen}{IRPAPERS} corpus. We leave it to future work to explore OCR transcription approaches such as Docling \cite{docling} and Olmocr \cite{olmocr}, which enable local inference and could reduce transcription costs, though potentially with different quality-speed tradeoffs.

From a developer experience perspective, these representations impose different kinds of friction. Text transcription introduces substantially higher computational and operational overhead. In practice, this overhead manifests in one of two ways. When using a hosted Multimodal Foundation Model API, transcription throughput is constrained by rate limits and pricing, resulting in long-running preprocessing pipelines whose cost scales linearly with corpus size. When self-hosting transcription models, the burden shifts to GPU provisioning, batching, fault tolerance, and service availability. In both cases, transcription introduces a non-trivial dependency on model capability and system reliability. In contrast, page-image preprocessing via base64 encoding is fast, deterministic, and easily parallelizable, requiring no external model inference dependencies. These tradeoffs extend beyond preprocessing into embedding and indexing costs, which we analyze in Section~3.

\section{Methodology}

\subsection{Text Retrieval}

Text search systems most commonly score query-document relevance using either sparse lexical representations or dense neural embeddings. We evaluate two approaches and their combination with hybrid search. BM25 \cite{bm25} is a sparse retrieval method that scores documents based on term frequency and inverse document frequency. BM25 excels at exact lexical matching, but is unable to capture semantic similarity between synonyms or related concepts. Dense retrieval encodes queries and documents into fixed-dimensional vectors using neural embedding models, then performs Maximum Inner Product Search (MIPS) to rank documents by embedding similarity. Hybrid search combines BM25 and dense retrieval scores to leverage both lexical precision and semantic understanding. We describe the fusion strategies used for hybrid text search in Section 3.4, where we also extend these methods to multimodal hybrid search.

\subsection{Image Retrieval with Late Interaction}

Retrieval models span a spectrum of representational choices, ranging from compact single-vector embeddings to more expressive late-interaction architectures that retain multiple token-level representations \cite{colbert}. Single-vector models offer simplicity and efficiency, and recent multimodal models demonstrate that they can achieve strong performance even on visually rich documents. Late-interaction methods, by contrast, preserve fine-grained alignment signals between queries and documents, enabling more precise matching at the cost of increased storage and computation. Late-interaction image retrieval has recently emerged as a promising approach for preserving fine-grained alignment between queries and visually rich documents. Rather than compressing an entire page into a single vector, these models retain multiple token- or patch-level embeddings and compute similarity via MaxSim aggregation, depicted in Figure 5 of Appendix C. MaxSim scoring enables precise matching between query terms and localized regions of a page. Several recent open-source works have explored the design space of these multi-vector image embedding models, such as ColPali, ColQwen2, and ColModernVBERT \cite{colpali, colmodernvbert}. ColPali and ColQwen2 pair large vision-language backbones with late-interaction scoring, achieving strong retrieval performance on benchmarks such as ViDoRe. However, both models are relatively large, with ColPali comprising approximately 2.9B parameters and ColQwen2 approximately 2.2B, resulting in substantial inference and storage costs. ColModernVBERT alternately combines a 150M-parameter bidirectional text encoder (ModernBERT) with a 100M-parameter vision encoder (SigLIP-2), for a total of 250M parameters. Despite being an order of magnitude smaller, ColModernVBERT achieves comparable performance on ViDoRe, 81.2 nDCG@5 versus 81.6 for ColPali, demonstrating that strong late-interaction image retrieval is possible with smaller models. We focus our primary analysis on ColModernVBERT due to this favorable performance–efficiency tradeoff. Its reduced parameter count enables faster inference, lower memory footprint, and more practical large-scale deployment, while still capturing the fine-grained alignment signals characteristic of late-interaction architectures. We include results for ColPali and ColQwen2 in Table 2 to contextualize performance across model sizes and design choices.

\subsection{Efficient Late Interaction with MUVERA}
A fundamental challenge with multi-vector embedding models and late interaction scoring is the storage of many vectors per document and the computational cost of MaxSim scoring. MUVERA \cite{muvera} addresses this computational expense by reducing multi-vector similarity search to standard single-vector MIPS with multi-vector rescoring. More particularly, this is achieved with the use of Fixed Dimensional Encoding (FDE), a transformation that converts a variable-sized set of embeddings into a single fixed-length vector. The approach partitions the embedding space into clusters using the SimHash locality-sensitive hashing algorithm. For each cluster, query embeddings falling into that region are summed, while document embeddings are averaged to form a centroid. The intuition is that if a query token and its best-matching document token land in the same cluster, their contribution to the similarity score is preserved. Multiple independent partitions are generated and concatenated to improve approximation quality, with configurable dimensionality reduction via random projections. With the use of FDEs, MUVERA is able to reuse the same HNSW, quantization, and rescoring implementations that have been developed in existing vector database systems.

The FDE transformation works as follows. First, each vector is assigned to a bucket using SimHash, a locality-sensitive hashing technique. SimHash draws $k_{sim}$ random hyperplanes through the embedding space. For each hyperplane, a vector receives a bit value of 1 or 0 depending on which side it falls. The resulting bit string determines the bucket assignment. For example, $k_{sim} = 4$ produces a 4-bit string yielding $2^4 = 16$ possible buckets. Because vectors that are close in space tend to fall on the same side of most hyperplanes, similar vectors land in the same bucket. Within each bucket, document vectors are averaged to form a representative sub-vector, while query vectors are summed. This is because in MaxSim scoring, each query token contributes independently to the final score, so multiple query tokens in the same bucket should all be counted rather than averaged away. Empty buckets are filled with the nearest available vector to reduce information loss. Each bucket's sub-vector is then compressed via random projection to $d_{proj}$ dimensions. Finally, all bucket sub-vectors are concatenated into a single FDE. Because the hyperplane assignments are random, the process is repeated multiple times with different random hyperplanes and the results concatenated. This yields the final FDE dimension: $2^{k_{sim}} \times d_{proj} \times \text{repetitions}$. For example, with $k_{sim}=4$, $d_{proj}=16$, and repetitions=10, each document is encoded as a single 2,560-dimensional vector regardless of its original multi-vector dimensionality.

MUVERA retrieval operates in two stages. First, at query time, the query's FDE is used to traverse the HNSW index using approximate similarity scores based on the compressed encodings. The HNSW search retrieves the top $ef$ candidate documents based on these approximate scores. Second, the original uncompressed multi-vector representations for these $ef$ candidates are fetched from storage and rescored using exact MaxSim computation to produce the final ranking. While $ef$ is a standard HNSW search parameter, in the MUVERA setting it additionally controls the number of approximate candidates that are subsequently rescored using exact MaxSim over the original multi-vector representations. Higher $ef$ values allow the system to reconsider more candidates with exact scoring. This two-stage approach enables standard HNSW indexing while reducing memory footprint compared to raw multi-vector storage. To quantify these memory savings, ColModernVBERT produces approximately 1,000 128-dimensional vectors per page, requiring $3{,}230 \times 1{,}000 \times 128 \times 4$ bytes = 1.65 GB for naive multi-vector storage. With MUVERA encoding using our parameters $k_{sim}=4$, $d_{proj}=16$, and repetitions=10, each page compresses to a single 2,560-dimensional vector, again with 4 bytes per dimension, this requires 33 MB to store 3,230 object vectors, a 50× reduction.

MUVERA fits within a broader line of work aimed at improving the efficiency and practicality of late-interaction retrieval models. Prior systems such as PLAID \cite{plaid} focus on execution-level optimizations for MaxSim scoring by leveraging centroid-based indexing and inverted lists. Query tokens are first matched against learned centroids to identify a reduced candidate set of documents, after which centroid-level approximations are used to filter candidates before residual decompression and exact MaxSim computation. This approach substantially reduces query-time computation by limiting where expensive late interaction is performed. In contrast, MUVERA changes how candidate generation is performed by approximating multi-vector similarity with fixed-dimensional encodings, enabling the use of standard single-vector ANN indexes while deferring exact MaxSim computation to a later rescoring stage. Orthogonal to both execution-level and query-time approximation techniques, recent work has explored reducing the storage and representation footprint of late-interaction models at indexing time. This includes token pooling approaches that cluster and average-pool token embeddings during index construction \cite{tokenpooling}, as well as learned token selection methods such as CRISP \cite{crisp}, which prune multi-vector representations by retaining only the most informative token embeddings. LEMUR \cite{lemur} proposes a learned reduction of late-interaction scoring, formulating MaxSim approximation as a supervised learning problem and producing compact single-vector representations that preserve late-interaction ranking behavior with substantially lower dimensionality. These lines of work suggest complementary axes for scaling late-interaction retrieval systems. We view the integration of such techniques with MUVERA-style approximations as a promising direction for future research.

\subsection{Multimodal Hybrid Search}
\label{sec:multimodal}

Given that text-based and image-based retrieval rely on different document representations, a natural question is whether combining them improves performance beyond either modality alone. Multimodal hybrid search addresses this by fusing scores from multiple retrieval systems. We combine three retrieval signals: BM25 sparse retrieval, Arctic 2.0 dense text embeddings, and ColModernVBERT multi-vector image embeddings. Combining these scores requires a fusion strategy. We evaluate two approaches. Relative Score Fusion (RSF) normalizes each retriever's scores to $[0,1]$ via min-max normalization, then computes a weighted sum. This preserves score magnitudes, allowing high-confidence predictions from any method to dominate the final ranking. Reciprocal Rank Fusion (RRF) assigns scores based on rank position, $\text{score}(d) = 1/(k + \text{rank}(d))$, then sums across methods, discarding score magnitudes entirely.

For hybrid text search, we fuse BM25 and Arctic 2.0 scores with equal weighting. For multimodal hybrid search, we fuse this hybrid text score with the ColModernVBERT image score using a parameter $\alpha \in [0,1]$, where $\alpha = 0$ uses text only and $\alpha = 1$ uses images only. At $\alpha = 0.5$, for example, the final score weights hybrid text and image retrieval equally, corresponding to effective weights of 0.25 for Arctic 2.0, 0.25 for BM25, and 0.5 for ColModernVBERT. Recent work has explored representation-level fusion as an alternative to output-level score combination. Guided Query Refinement from Uzan et al.~\cite{GQR} iteratively adjusts query embeddings using gradient descent guided by a complementary retriever. We leave evaluation of representation-level fusion to future work.

\subsection{Question Answering with Retrieval-Augmented Generation (RAG)}

Retrieval-Augmented Generation (RAG) combines document retrieval with a downstream reader model to answer questions using external evidence. Given a query, a RAG system first retrieves the top-k most relevant documents from a corpus, then conditions a language model on this retrieved context to generate an answer. In the context of scientific literature, RAG is particularly well-suited to questions that require precise methodological details, numerical results, or evidence grounded in specific sections of a paper. The effectiveness of a RAG system therefore depends not only on the reader model, but critically on the relevance of the retrieved documents. We evaluate two RAG configurations that differ in the modality of retrieved context. ImageRAG retrieves page images using ColModernVBERT and provides these images directly to a multimodal reader model, preserving layout, figures, and visual structure. TextRAG retrieves OCR-transcribed text using hybrid text retrieval and provides the resulting text to the same reader model. Both configurations use identical generation prompts and evaluation protocols, isolating the impact of retrieval modality and depth on downstream question answering performance.

\subsection{Experimental Methodology}

We report Recall@K for our retrieval analysis, also known as Success@K in single-relevance settings \cite{colbertv2}. This contrasts with benchmarks like ViDoRe V3 \cite{vidorev3}, which report NDCG@10 to capture ranking quality across their multi-relevance annotations where queries average 5.1 relevant documents. We encode PDF page images with ColModernVBERT multi-vector image embeddings, producing an average of 1,000 128-dimensional vectors per page. We encode text transcriptions with Arctic 2.0, producing 1,024-dimensional embeddings per transcript \cite{arctic-embed}. We use Weaviate's implementation of relative score fusion to combine BM25 with these Arctic 2.0 dense embeddings for our hybrid text search retrieval method \cite{weaviate}. For multimodal hybrid search, we first compute hybrid text scores using fixed equal weighting of Arctic 2.0 and BM25, then fuse with ColModernVBERT image scores using parameter $\alpha \in [0, 1]$. Thus $\alpha$ = 0.5 corresponds to effective weights of 0.25 (Arctic 2.0), 0.25 (BM25), and 0.5 (ColModernVBERT).  We use the Weaviate Database for HNSW search, BM25, hybrid fusion, and MUVERA, as well as embedding inference for ColModernVBERT and Arctic 2.0 \cite{weaviate}. All HNSW-based retrieval systems with single vector representations use a fixed search parameter of $ef$ = 160. We further evaluate retrieval-augmented generation (RAG) systems that combine document retrieval with a reader model to answer questions. Given a query, the system first retrieves the top-$k$ most relevant documents, then passes them as context to a large language model that generates an answer. We implement two RAG systems with DSPy \cite{dspy}, to enable comparisons between retrieval strategies and reader modality, ImageRAG and TextRAG. ImageRAG uses ColModernVBERT to retrieve relevant page images, which are passed directly to GPT-4.1 as visual context. This preserves the original document layout including figures, tables, and formatting. TextRAG uses hybrid text search to retrieve relevant text transcriptions, which are passed to GPT-4.1 as textual context. We evaluate answer quality using an LLM-as-Judge protocol. For each question, the judge receives the tuple (question, system\_answer, reference\_answer) and returns a binary score indicating whether the system answer is semantically equivalent to the reference answer (Appendix~\ref{app:alignment_score_prompt}). To reduce judgment variance, we run the judge three times with independent sampling and take the majority vote. The reported alignment score is the fraction of questions scored True.

\section{Experimental Results}

\subsection{Retrieval Results}

\begin{figure}
    \centering
    \includegraphics[width=1.0\linewidth]{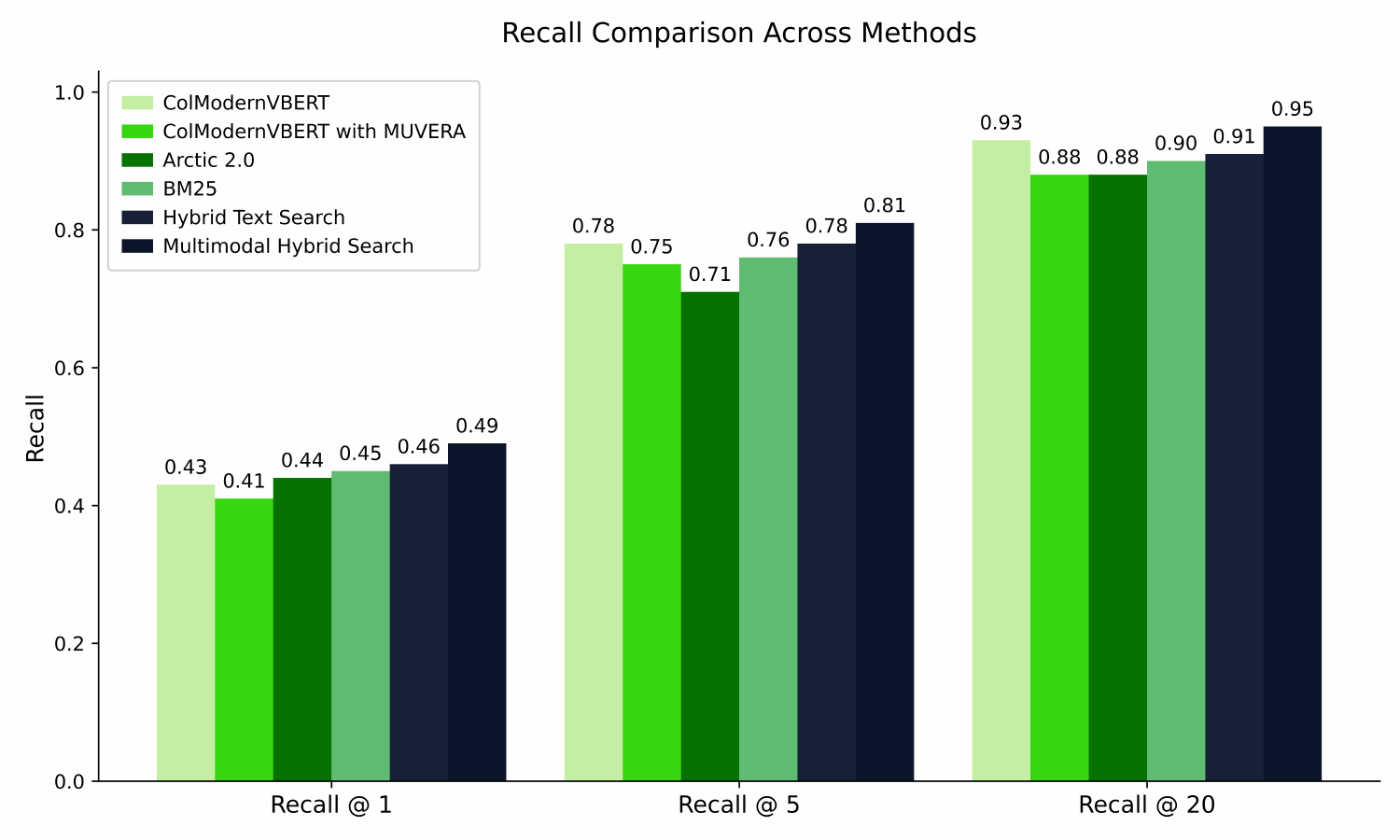}
    \caption{Results of our retrieval test. Multimodal hybrid search consistently outperforms single-modality retrieval across all recall levels, highlighting the complementary strengths of text and image representations.}
    \label{fig:placeholder}
\end{figure}

Figure 2 presents our results comparing text- and image-based retrieval with open-source models. We find that text-based retrieval provides stronger signals for identifying the single best-ranked document, while image-based retrieval is comparably effective at recovering relevant pages when retrieval depth increases. Hybrid text search, combining BM25 and Arctic 2.0 embeddings using relative score fusion (RSF) with $\alpha$ = 0.5, achieves the strongest open-source text performance at 46\% Recall@1, while ColModernVBERT multi-vector image retrieval reaches 43\% Recall@1. At deeper recall levels, image retrieval achieves 93\% Recall@20, exceeding hybrid text search at 91\%. These two approaches exhibit complementary failure modes. At Recall@1, 22 queries succeed with text, but fail with images, while 18 succeed with images, but fail with text. This result motivates the use of multimodal hybrid search, which fuses both signals to exceed single-modality performance. Figure 3 presents a hyperparameter sweep of $\alpha$ values for weighting results with reciprocal rank fusion (RRF) and relative score fusion (RSF). Multimodal hybrid search with RSF achieves 49\% Recall@1, a 3-point absolute improvement over hybrid text search alone. At Recall@5, multimodal hybrid search achieves a 3-point absolute improvement over the next highest performing open-source method, and a 2-point absolute improvement at Recall@20. By exploiting complementary retrieval failures, multimodal fusion breaks the performance ceiling imposed by either modality alone.

We further compare multi-vector image embedding models in Table 2. ColPali and ColQwen2 achieve higher Recall@1 than ColModernVBERT, with ColQwen2 reaching 49\%, while all models achieve similar Recall@20 in the 93–94\% range. This pattern suggests that increased model capacity primarily improves top-rank discrimination. Given the substantial differences in parameter count used in ColModernVBERT compared to ColPali or ColQwen2, the results highlight saturating returns with the current state of larger multi-vector image embedding models. While multi-vector image embeddings provide strong retrieval performance, they introduce significant storage and computation overhead. MUVERA encoding addresses this challenge by compressing multi-vector representations into fixed-dimensional encodings suitable for approximate nearest neighbor search. Table 3 quantifies the retrieval impact of this compression. Relative to exact multi-vector retrieval, MUVERA incurs a 2-point drop at Recall@1 and a 5-point drop at Recall@20 when $ef$ is set to 1024. Reducing ef exacerbates this effect, demonstrating how MUVERA can be tuned to achieve a desired tradeoff between retrieval performance and system efficiency.

We conclude our analysis with an evaluation of closed-source embedding models to calibrate the open-source results against current proprietary capabilities. Shown in Table 4, Cohere Embed v4 image embeddings achieve 58\% Recall@1, substantially exceeding open-source image and text baselines, while Voyage 3 Large text embeddings reach 52\% Recall@1. The performance gap narrows at deeper recall levels. Cohere v4 exceeds the strongest open-source image model by 3 points at Recall@20. Multimodal fusion of closed-source models further improves Recall@5 and Recall@20, reaching 91\% and 98\%, respectively, but does not improve Recall@1 beyond Cohere Embed v4 alone. At Recall@1, Cohere Embed v4 exclusively succeeds on 25 queries, while Voyage 3 Large exclusively succeeds on 15 queries. More detailed results of our hyperparameter sweep of multimodal hybrid search $\alpha$ values for Cohere Embed v4 and Voyage 3 Large can be found in Figure 4 of Appendix B.

Taken together, these results show that retrieval performance in scientific document corpora is governed by how representational choices, model capacity, and system constraints interact across ranking depths. Among open-source systems, text-based retrieval provides stronger top-ranked precision, while image-based retrieval recovers a comparable or larger set of relevant pages at deeper recall, and multimodal hybrid search exploits their complementary failure modes to exceed single-modality ceilings. Increasing image model capacity, from ColModernVBERT to ColPali and ColQwen2, primarily improves top-rank discrimination while yielding diminishing returns at higher recall, indicating that deeper recall stabilizes once the relevant neighborhood is reliably retrieved. MUVERA exposes a controllable performance-efficiency tradeoff through its two-stage retrieval design, where the first stage HNSW search parameter $ef$ determines how many approximate candidates are subsequently rescored using exact MaxSim. While all non-MUVERA baselines use a fixed HNSW search parameter of $ef$ = 160, increasing $ef$ in MUVERA expands the candidate set subjected to exact MaxSim rescoring, allowing partial recovery of fine-grained matching signals lost with FDE compression. Finally, closed-source image embeddings achieving the strongest top-ranked performance with narrowing gaps at higher recall levels.

\begin{figure}
    \centering
    \includegraphics[width=1.0\linewidth]{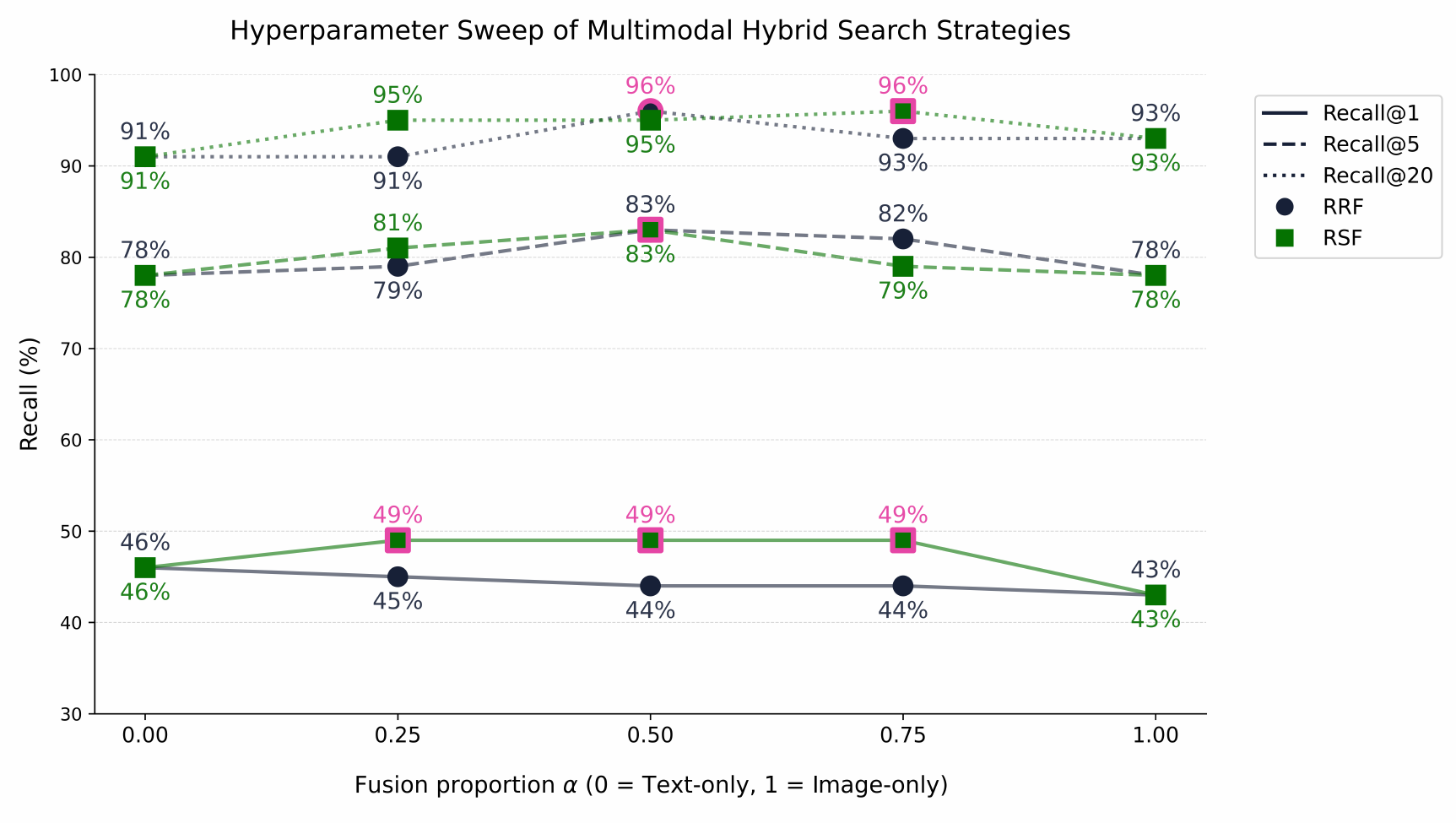}
    \caption{Comparison of Multimodal Hybrid Search fusion strategies combining hybrid text search (Arctic 2.0 + BM25) with ColModernVBERT image embeddings. For each value of $\alpha$, we report results for both RRF and RSF, visualized as distinct fusion strategies, where $\alpha$=0 uses text only and $\alpha$=1 uses images only. The highest performing configuration(s) for each respective recall target (@1, @5, and @20) are highlighted in pink.}
    \label{fig:placeholder}
\end{figure}

\begin{table}[h]
\centering
\label{tab:retrieval}
\setlength{\tabcolsep}{12pt}
\renewcommand{\arraystretch}{1.5}
\begin{tabular}{|l|c|c|c|}
\hline
\textbf{Retriever} & \textbf{Recall@1} & \textbf{Recall@5} & \textbf{Recall@20} \\
\hline
ColModernVBERT & 43\% & 78\% & 93\% \\
\hline
ColPali & 45\% & 79\% & 93\% \\
\hline
ColQwen2 & \textbf{49\%} & \textbf{81\%} & \textbf{94\%} \\
\hline
\end{tabular}
\vspace{1em}
\caption{Comparison of the ColModernVBERT, ColPali, and ColQwen2 Multi-Vector Image Embedding Models.}
\end{table}

\begin{table}[h]
\centering
\label{tab:muvera}
\setlength{\tabcolsep}{12pt}
\renewcommand{\arraystretch}{1.5}
\begin{tabular}{|l|c|c|c|}
\hline
\textbf{Configuration} & \textbf{Recall@1} & \textbf{Recall@5} & \textbf{Recall@20} \\
\hline
ColModernVBERT (No MUVERA) & \textbf{43\%} & \textbf{78\%} & \textbf{93\%} \\
\hline
MUVERA ($ef$=1024) & 41\% & 75\% & 88\% \\
\hline
MUVERA ($ef$=512) & 37\% & 68\% & 78\% \\
\hline
MUVERA ($ef$=256) & 35\% & 61\% & 66\% \\
\hline
\end{tabular}
\vspace{1em}
\caption{Information loss analysis of MUVERA encoding and rescoring across values of $ef$. All single vector retrieval methods use HNSW with $ef$ = 160.}
\end{table}

\begin{table}[h]
\centering
\label{tab:retrieval}
\setlength{\tabcolsep}{12pt}
\renewcommand{\arraystretch}{1.5}
\begin{tabular}{|l|c|c|c|}
\hline
\textbf{Retriever} & \textbf{Recall@1} & \textbf{Recall@5} & \textbf{Recall@20} \\
\hline
Cohere Embed v4.0 (Images) & \textbf{58\%} & 87\% & 97\% \\
\hline
Voyage 3 Large (Text) & 52\% & 86\% & 95\% \\
\hline
Multimodal Hybrid Search & \textbf{58\%} & \textbf{91\%} & \textbf{98\%} \\
\hline
\end{tabular}
\vspace{1em}
\caption{Results of Closed-Source Cohere Embed 4.0 image embeddings and Voyage 3 Large text embeddings, as well as their combination in multimodal hybrid search.}
\end{table}

\subsection{Question Answering Results}

Our question answering experiments reveal the importance of both retrieval quality and context quantity for RAG performance. We begin by evaluating three baseline conditions to validate task difficulty: No Retrieval, Hard Negative, and Oracle Retrieval. No Retrieval provides only the question to the reader model, testing whether answers can be derived from parametric knowledge alone. Hard Negative retrieval provides the top-ranked document excluding the gold page as context, testing whether similar, but incorrect, documents suffice for both text and image modalities. For text-based systems, this context is provided as OCR-transcribed text, while for image-based variants, the corresponding page image is provided to the multimodal reader model. Oracle Retrieval bypasses the retriever and directly provides the known ground-truth page as context, establishing a reference point for single-document question answering with perfect retrieval. Oracle retrieval follows the same modality-specific protocol, providing either the ground-truth page image in ImageRAG or its OCR transcription in TextRAG. The No Retrieval baseline achieves only 0.16 alignment, establishing that \textcolor{darkgreen}{IRPAPERS} questions cannot be answered from parametric knowledge alone. Hard Negative TextRAG achieves 0.39 alignment and Hard Negative ImageRAG achieves 0.12. With standard retrieval at k=1, TextRAG achieves 0.62 alignment compared to 0.40 for ImageRAG. Both methods require about 1,300 input tokens per question on average, with text requiring roughly 11\% more input tokens than images. Oracle retrieval at k=1 improves these results to 0.74 and 0.68 respectively. Increasing retrieval depth to k = 5 yields the strongest results. TextRAG achieves 0.82 alignment and ImageRAG achieves 0.71. This comes at increased token cost proportional to k, with input tokens rising from 1,366 to 6,022 for text representations and from 1,228 to 5,200 for images. Interestingly, k = 5 outperforms an oracle k = 1 for both modalities, suggesting that additional context from related pages provides valuable supporting information for answer synthesis even when those pages are not the exact gold source. However, ImageRAG suffers a steeper decline when reducing k from 5 to 1 (0.71 to 0.40) compared to TextRAG (0.82 to 0.62), indicating that image-based systems benefit from higher values of K for question answering more than text-based systems. We leave it to future work to evaluate whether this gap persists across different reader models to determine whether our findings reflect fundamental modality differences or model-specific behaviors.

\begin{table}[h]
\centering
\label{tab:qa}
\setlength{\tabcolsep}{12pt}
\renewcommand{\arraystretch}{1.5}
\begin{tabular}{|l|c|c|c|}
\hline
\textbf{System} & \textbf{Alignment Score} & \textbf{Avg Input Tokens} & \textbf{Avg Output Tokens} \\
\hline
No Retrieval Baseline & 0.16 & 173 & 135 \\
\hline
Hard Negative Image Context $(k = 1)$ & 0.12 & 1233 & 134 \\
\hline
Hard Negative Text Context $(k = 1)$ & 0.39 & 1304 & 162 \\
\hline
Oracle Image Retrieval $(k = 1)$ & 0.68 & 1208 & 125 \\
\hline
Oracle Text Retrieval $(k = 1)$ & 0.74 & 1294 & 155 \\
\hline
ImageRAG $(k = 1)$ & 0.40 & 1228 & 124 \\
\hline
TextRAG $(k = 1)$ & 0.62 & 1366 & 160 \\
\hline
ImageRAG $(k = 5)$ & 0.71 & 5200 & 178 \\
\hline
TextRAG $(k = 5)$ & \textbf{0.82} & 6022 & 243 \\
\hline
\end{tabular}
\vspace{1em}
\caption{Results of our question answering test. All image-based systems provide page images as input, and all text-based systems provide OCR transcribed text.}
\end{table}

\section{Limitations of Unimodal Representations}
Multimodal hybrid search outperforms either modality in isolation, yet this aggregate improvement masks a more nuanced finding; text and image representations each succeed on queries where the other fails. At Recall@1, image-based retrieval successfully retrieves relevant documents for 18 queries that text-based retrieval misses, while text-based retrieval succeeds on 22 queries that image-based retrieval fails to retrieve. This asymmetry persists across different unimodal model pairings. Cohere Embed v4 image embeddings exclusively succeeds on 25 queries where Voyage 3 Large text embeddings exclusively succeeds on 15 queries. These complementary successes and failures indicate that neither modality is redundant. Each modality captures relevance signals that are not being recovered from the other. These results motivate a deeper investigation into the limitations of unimodal representations. Rather than asking which modality is better in aggregate, we seek to understand why each modality fails, what kinds of information those failures correspond to, and whether those failures arise from fundamental representational constraints.

To better understand the role of image representations, we first examine whether visual grounding is necessary for answering questions derived from figures in scientific papers. Restricting our analysis to the 19 Query Writing papers used throughout \textcolor{darkgreen}{IRPAPERS}, we identify 63 visual elements and manually categorize them into data charts (32), architectural diagrams (10), and conceptual or abstract visuals (21). Using a targeted prompt that asks an LLM to generate questions that should be impossible to answer using the OCR text transcription alone (Appendix A.5), we derive 30 visual questions. For the remaining visuals, the LLM explicitly responds that it cannot imagine a question uniquely answerable from the visual content. On this adversarial set under oracle retrieval conditions, text-based question answering still outperforms image-based question answering, achieving an alignment score of 0.67 compared to 0.53. This result reflects a broader property of scientific communication, most figures encode information that is either explicitly described in accompanying prose or can be faithfully serialized into structured text through OCR. Authors typically describe figure content in captions or surrounding text, and OCR transcription captures tabular and diagrammatic structure with sufficient fidelity for our question-answering task.

This conclusion does not hold for all visual elements. To isolate cases where text representations fundamentally fail, we construct a focused adversarial study centered on a single abstract visualization in the corpus: a t-SNE plot depicting the relative clustering of query rewrites in an embedding space. Unlike charts or architectural diagrams, the informational content of this visualization depends on relative positioning and cluster geometry, properties that are difficult to articulate precisely in text. Text transcriptions of t-SNE plots typically include only axis labels and legends, omitting the spatial coordinates and cluster boundaries that constitute the primary informational content. Using the same prompt from Appendix A.5, we derive ten questions from this figure, samples of which are shown in Table 8. Under oracle retrieval conditions, image-based QA achieves 70\% accuracy while text-based QA drops to 30\%. While this result is limited in scope, it raises a broader question: \textit{Are there additional examples of scientific visuals that cannot be easily captured in text?} Candidate examples include manifold visualizations in machine learning, vector field diagrams in physics, or geometric depictions of latent spaces. What unites these examples is that answering such questions often requires interpreting geometric properties that are difficult to convey fully through symbolic descriptions. The t-SNE visual, as well as sample questions derived from it, can be found in Appendix E.

Image representations, however, exhibit complementary limitations. While they preserve visual structure, they provide no direct mechanism for enforcing exact textual constraints. Recent advances in structured outputs for large language models have enabled agentic retrieval systems that can apply explicit constraints to search queries, such as exact-match filters or schema-guided routing \cite{structuredrag, queryingdbswithfc}. For instance, a query asking "What does HyDE stand for?" requires retrieving pages containing the exact string "HyDE." Visually similar pages discussing other dense retrieval methods may rank highly under image embeddings despite lacking the target acronym entirely. This pattern explains why BM25 contributes meaningfully to hybrid text search. It provides exact lexical matching that dense semantic embeddings do not explicitly encode. There is no clear analogue in image-based retrieval to exact string presence or symbolic matching, highlighting a fundamental asymmetry in how symbolic information is accessed across modalities.

The visual question analysis presented here is necessarily limited by dataset scale and by the rarity of abstract visualizations in our curated domain of scientific papers. Notably, fewer than half of the identified visual elements yield questions that an LLM deems uniquely visual, underscoring how uncommon such visualizations appear in information retrieval research. Future work could leverage prompt optimization techniques for question generation to more systematically elicit questions that target representational blind spots \cite{UDAPDR}, or expand evaluation to corpora richer in geometric and spatial visualizations. Beyond opportunities to improve our adversarial evaluation methodology, these findings point toward practical system design. Our results motivate two opportunities for future work integrating an agentic layer with multimodal hybrid search. First, agentic systems could dynamically adjust the weighting parameter, $\alpha$, based on query characteristics, emphasizing image signals for visually grounded information and text signals for symbolic precision \cite{routerretriever, freshstack}. Second, at question-answering time, images need only be passed to the reader model when the query explicitly targets information that cannot be resolved from text alone. This decouples the role of images in retrieval scoring from their role in answer generation. Image embeddings can contribute to relevance ranking while the images themselves remain on disk, retrieved and passed to the reader only when visual grounding is required. Together, these strategies allow systems to exploit the complementary strengths of each representation while minimizing unnecessary computational and token overhead.

\section{Related Works}

\subsection{Visual Document Benchmarks}

Visual document understanding has been evaluated across several complementary benchmarks. ViDoRe v3 \cite{vidorev3, vidorev2} represents the current gold standard for enterprise visual document retrieval evaluation. The benchmark contains 26,000 pages across 10 industrial domains (finance, pharmaceuticals, HR, energy, and others) with about 1,000 to 5,000 pages per domain, along with 3,099 human-annotated queries labeled across 7 categories including multi-hop, numerical, open-ended, and more. The authors evaluate several visual document retrieval models, finding that nemo-colembed-3b \cite{nemocolembed} achieves the strongest performance at 65.6\% NDCG@10 on English datasets, followed by nemo-colembed-1b (64.3\%) and jina-v4 (63.9\%) \cite{jinav4}. ColModernVBERT, the model we primarily focus on in our analysis, achieves 50.7\% nDCG@10. ViDoRe V3 reports NDCG@10 to capture ranking quality across their multi-relevance annotations where queries average 5.1 relevant pages, whereas \textcolor{darkgreen}{IRPAPERS} reports Recall@K because our needle-in-the-haystack methodology assigns a single gold page per query. \textcolor{darkgreen}{IRPAPERS} complements ViDoRe V3 by providing depth within a single research community: our 166 papers and 3,230 pages could constitute an "Academic IR" domain comparable in scale to individual ViDoRe V3 subsets, while our queries specifically test whether retrieval systems can discriminate among papers sharing vocabulary, methods, and subject matter, a challenge distinct from cross-domain enterprise retrieval.

FinanceBench \cite{financebench} is a benchmark for open-book financial question answering comprising 10,231 questions about publicly traded companies with corresponding answers and evidence strings extracted from 10-Ks, 10-Qs, 8-Ks, and earnings reports. The benchmark emphasizes ecological validity, with questions designed through interviews with financial analysts and categorized into information extraction, numerical reasoning, and logical reasoning types. Their evaluation of 16 model configurations revealed that GPT-4-Turbo with a shared vector store incorrectly answered or refused to answer 81\% of questions, while even the best-performing long-context configuration achieved only 79\% accuracy. Critically, FinanceBench treats documents exclusively as text relying on standard PDF text extraction—and does not explore visual document representations. In contrast, \textcolor{darkgreen}{IRPAPERS} directly investigates whether preserving the visual structure of PDF pages through image embeddings can match or exceed text-based retrieval. Furthermore, while FinanceBench evaluates traditional dense embeddings and BM25, we extend the evaluation to late-interaction multi-vector methods and MUVERA encoding.

Recent work has begun to explicitly evaluate the ability of vision–language models to read and reason over dense textual content presented visually. ReadBench \cite{ReadBench} introduces a multimodal benchmark that converts established text-only benchmarks into images of text, enabling controlled comparisons between textual and visual reading settings. Its results show that while performance degradation is modest for short inputs, accuracy drops sharply for long, multi-page visual contexts, highlighting persistent limitations in reasoning over visually presented text that are not captured by traditional OCR-focused or visual understanding benchmarks. DocVQA \cite{docvqa} introduced a benchmark for Visual Question Answering on document images, consisting of 50,000 questions defined on over 12,000 document images from diverse document types including forms, invoices, and reports. The benchmark emphasizes that understanding documents requires going beyond OCR to comprehend layout, structure, and non-textual elements, with existing models showing a large performance gap compared to human performance (94.36\% accuracy) particularly on questions requiring structural document understanding. However, DocVQA primarily evaluates single-document comprehension rather than retrieval. Questions such as, \textit{What date is seen on the seal at the top of the letter?} or \textit{What is the ZIP code written?} presuppose that the correct document has already been identified. These document-specific questions do not naturally extend to retrieval benchmarks, where queries must discriminate among candidate documents without prior knowledge of which document contains the answer. Furthermore, DocVQA's corpus spans diverse document types, invoices, memos, letters, reports, whereas \textcolor{darkgreen}{IRPAPERS} focuses solely on scientific papers. SlideVQA \cite{SlideVQA} and InfographicVQA \cite{InfographicVQA} extend visual document understanding to presentation slides and infographics respectively, with SlideVQA notably requiring multi-hop reasoning across multiple pages. However, like DocVQA, both benchmarks evaluate comprehension over provided documents rather than retrieval from a corpus, and neither addresses the challenge of discriminating among semantically similar documents within a single research domain.

\subsection{Scientific Literature Mining}

Our work aims to build on recent works exploring AI systems for scientific literature. Large-scale scientific corpora such as S2ORC \cite{s2orc} and AI2's Paper Finder \cite{paperfinder} have enabled significant advances in scholarly information access. Most similarly to \textcolor{darkgreen}{IRPAPERS}, Ajith et al. introduce LitSearch \cite{litsearch}. LitSearch proposes a benchmark for scientific literature search designed to reflect how researchers find prior work. LitSearch generates information-seeking questions by rewriting inline citation contexts into standalone search queries using LLMs, followed by filtering to reduce lexical overlap with target papers and manual validation to ensure realism and discriminative difficulty. In contrast to \textcolor{darkgreen}{IRPAPERS}, which concentrates on page-level retrieval and question answering within a tightly scoped Information Retrieval corpus using both image and OCR-based representations, LitSearch operates over a broader ML and NLP literature and evaluates document-level retrieval using titles and abstracts alone. Their experimental results show large gains from instruction-tuned dense retrievers over BM25, modest but consistent improvements from LLM-based reranking, and limited benefits from full-text representations relative to abstracts. Interestingly, LitSearch reports that state-of-the-art dense retrievers outperform widely used scientific search engines such as Google Scholar, underscoring a gap between general-purpose scholarly search systems and task-specific retrieval models. Together, these findings complement our results by highlighting how benchmark design, question construction, and representational scope fundamentally shape conclusions about retrieval effectiveness in scientific literature mining.

Press et al. introduce CiteME \cite{CiteME}, a manually curated citation attribution benchmark that evaluates whether language models can correctly identify a cited paper given only a text excerpt with an anonymized citation placeholder. Their benchmark consists of 130 carefully selected excerpts from recent ML papers, each unambiguously referencing a single paper, with instances that GPT-4o could answer from memorization filtered out. In contrast to \textcolor{darkgreen}{IRPAPERS}, which evaluates page-level retrieval and question answering over a document corpus using both image and OCR-based representations, CiteME frames citation attribution as an open-ended retrieval task where models must identify papers across the entire Semantic Scholar index of 218 million papers using only the semantic content of a text excerpt. Their experimental results reveal a substantial gap between human performance (69.7\% accuracy in under 40 seconds) and frontier LMs (4.2–18.5\% without tools). The authors further introduce CiteAgent, an autonomous system using GPT-4o with Semantic Scholar search capabilities, and achieve 35.3\% accuracy. Notably, previous state-of-the-art retrieval systems (SPECTER and SPECTER2 \cite{SPECTER, SPECTER2} achieved 0\% accuracy on CiteME, and the authors identify three distinct error categories: misunderstanding the excerpt (50\%), stopping at approximately matching papers (32\%), and finding the correct citation in references but selecting the citing paper instead (18\%). Together, these findings complement our results by highlighting how the granularity of retrieval targets, page-level documents versus open-domain paper identification, and the nature of the query signal, information-seeking questions versus citation contexts, shape methodology and conclusions about retrieval effectiveness in scientific literature mining.

Prior scientific QA benchmarks designed for agent training focus on single-hop factoid questions to enable verifiable reward signals \cite{PaperSearchQA}. PaperSearchQA introduces a large-scale environment for training search agents over 16 million biomedical abstracts using reinforcement learning with outcome-only supervision. Questions are synthetically generated from abstracts to ensure unambiguous, single-document grounding. The authors find that training agents to search in this environment with reinforcement learning yields substantial gains over non-RL baselines, 9.6 and 5.5 point gains on PaperSearchQA and BioASQ , respectively, for 3B parameter LLMs, while for 7B parameter LLMs these gains increase to 14.5 and 9.3 points. Together, these benchmarks highlight complementary dimensions of scientific literature mining: how agents learn to search, and what information is preserved or lost under different document representations.

Several works have further expanded the landscape of AI in Scientific Literature Mining toward higher-level tasks such as fact verification, hypothesis generation, and autonomous research agents. These tasks introduce substantially weaker grounding signals and rely heavily on expert judgment for evaluation. Even more grounded tasks remain incompletely understood. Existing scientific figure benchmarks such as ChartQA \cite{chartqa}, SciGraphQA \cite{scigraphqa}, and SPIQA \cite{spiqa} evaluate visual comprehension rather than retrieval, leaving open the question of how page-image representations perform when the objective is to find relevant scientific documents. DeepScholar-Bench \cite{deepscholarbench} further explores this boundary by evaluating deep, document-grounded scholarly reasoning that requires integrating evidence across entire research papers. Unlike retrieval-centric benchmarks such as \textcolor{darkgreen}{IRPAPERS}, its evaluation relies on expert judgment of answer quality rather than objective relevance metrics, highlighting how evaluation becomes more subjective even when grounding remains strong. In a large-scale human study, Si et al. \cite{canllmsgenerate} ask \textit{Can LLMs Generate Novel Research Ideas?} The authors show that while LLMs can produce ideas judged as more novel than those of expert researchers, progress on such weakly grounded tasks is dominated by evaluation challenges, including low inter-reviewer agreement and the unreliability of LLM-based judges. Recent systems such as The AI Scientist \cite{theaiscientist} extend this direction by automating end-to-end research workflows from hypothesis generation to experiment execution and paper drafting. This work depends on task-specific heuristics and human validation to assess scientific quality and novelty. In contrast, benchmarks such as RE-Bench \cite{rebench} and emerging frameworks for the Automated Design of Agentic Systems (ADAS) \cite{adas} emphasize tightly grounded, executable environments with objective scoring functions and direct human baselines, demonstrating that stronger grounding can mitigate some evaluation fragility. Taken together, these works underscore that as scientific tasks move beyond retrieval and question answering toward open-ended discovery and autonomous experimentation, evaluation methodology becomes a central challenge and a key determinant of progress.

\section{Conclusion}

We introduced \textcolor{darkgreen}{IRPAPERS}, a benchmark for evaluating visual document retrieval and question answering over scientific literature, comprising 166 information retrieval papers (3,230 pages) with 180 curated questions targeting precise methodological details. Our systematic comparison reveals that text-based and image-based approaches exhibit complementary strengths. Hybrid text search achieves 46\% Recall@1 compared to 43\% for ColModernVBERT image embeddings, yet image retrieval matches or exceeds text at deeper recall levels. Critically, 22 queries succeed exclusively with text while 18 succeed exclusively with images, enabling multimodal hybrid search to reach 49\% Recall@1 and 95\% Recall@20, surpassing either modality alone. Larger multi-vector image embedding models improve top-rank performance, ColPali reaches 45\% and ColQwen2 reaches 49\% Recall@1, while converging at Recall@20 (93–94\%), indicating diminishing returns from increased model capacity at deeper ranking depths. MUVERA encoding achieves 41\% Recall@1 with $ef$ = 1024 and 35\% at $ef$ = 256, illustrating a tunable tradeoff between efficiency and quality. Among closed-source models, Cohere Embed v4 achieves 58\% Recall@1 on page images, outperforming Voyage 3 Large text embeddings at 52\% Recall@1 and establishing a 9-point gap over the best open-source approach. For question answering, text-based RAG outperforms image-based RAG (0.82 vs. 0.71 alignment at k=5), though both benefit substantially from increased retrieval depth, with five retrieved documents outperforming oracle single-document retrieval.

These findings point toward several directions for future work. The complementary failure modes between modalities motivate adaptive multimodal systems that dynamically weight text and image signals based on query characteristics. Additionally, while text transcriptions suffice for most scientific content, abstract visualizations like t-SNE plots encode spatial relationships that resist textual description, whereas image representations lack mechanisms for exact lexical matching. Agentic search architectures that combine semantic retrieval with structured constraints could bridge this gap, addressing the precision requirements of scientific literature mining. \textcolor{darkgreen}{IRPAPERS} provides a foundation for systematically understanding how text- and image-based representations perform on retrieval and question answering with scientific papers.

\bibliographystyle{plain}
\bibliography{references}

\appendix

\section{Prompts}
\label{app:prompts}

\subsection{PDF Transcription Prompt}
\label{app:transcription_prompt}

The following prompt was used to transcribe PDF pages into structured text using OpenAI's GPT-4.1:

\begin{quote}
Can you please transcribe the content from this PDF page into a text format?
If you come across tables, please transcribe them into markdown formatted text.
If you come across images, please just omit them and use their provided caption in the text format.
Please do not begin your response with ``Here is the transcription of the image:'' or anything similar, just start with the text transcription.
\end{quote}

\subsection{Query and Answer Generation Prompt}
\label{app:query_prompt}

The following prompt was used with Claude Sonnet 4.5 to generate questions and answers for each page:

\begin{quote}
I need your help simulating information seeking questions that information retrieval (IR) scientists would ask about the paper attached to this message.

Could you please write a question that each page of the attached paper uniquely answers? Further, these questions will be used to find this particular page amongst a large collection of other research papers, therefore, please try to make the question highly specific to this particular paper. Don't make the question too complicated, it should have a relatively short answer, not more than one or two sentences.

Please further explain why this question is only answerable by the cited page and not others, and why this question is only answerable by this particular paper and not other IR papers. Please also provide the answer to the question.

Please follow these instructions when generating these questions, answers, and explanations:

Avoid phrases such as ``this study'', make sure the questions are self-contained as if you are searching through a large collection of similar papers for the answers to these questions. Please use the specific name of the paper if you would like to make these kind of references, but generally please avoid doing so.

Avoid the use of ``the researchers'', if referencing a particular study please use the name of the paper.

Do not reference the study as ``the paper'', use the name of the paper always. Remember, these questions must be self-contained.

Avoid the use of mathematical equations in your answers.

Remember, do not overcomplicate the question, it should have a fairly short answer, not more than one or two sentences.
\end{quote}

\subsection{RAG Prompt}
\label{app:rag_prompt}
The following DSPy Signature \cite{dspy} was used to answer questions in our RAG question answering tests.

\begin{lstlisting}[style=python]
class GenerateAnswerFromParameters(dspy.Signature):
    """Answer the question as well as you can."""
    question: str = dspy.InputField(description="The question to answer.")
    answer: str = dspy.OutputField(description="The answer to the question.")
\end{lstlisting}

\subsection{Alignment Score Prompt}
\label{app:alignment_score_prompt}
The following DSPy Signature \cite{dspy} was used to assess the alignment between ground truth and system answers. The judge outputs a boolean score. We define the alignment score for a system as the mean of this boolean across all questions. We run the judge three times per question and take the majority vote before averaging.

\begin{lstlisting}[style=python]
class AssessAlignmentScore(dspy.Signature):
    """You are an expert grader assessing if a system's answer is 
    semantically aligned with the correct answer.
    Only return True if the system answer has essentially the same 
    meaning as the correct answer.
    If the system answer misses key aspects or meaning, return False.
    """
    question: str = dspy.InputField(description="The question asked.")
    system_answer: str = dspy.InputField(
        description="The answer generated by the system.")
    correct_answer: str = dspy.InputField(
        description="The reference answer containing the 
        correct and complete information."
    )
    score: bool = dspy.OutputField(
        description="True if system_answer is equivalent in 
        meaning to correct_answer, otherwise False."
    )
\end{lstlisting}

\subsection{Adversarial Text Representation Question Generation Prompt}
\label{app:query_prompt}

The following prompt was used with Claude Sonnet 4.5 to generate questions and answers for each page intended to adversarially attack using an OCR text transcription to represent the page instead of its image.

\begin{quote}
I am researching the difference between storing PDF papers as images or OCR text transcriptions for modern AI systems. I need your help finding questions that highlight the unique strengths of having access to the image representation of the PDF for question answering.

Are there any visuals in this paper that answer questions that would be impossible to answer with just an OCR to text representation of the paper?

These questions shouldn’t be very specific details of the image, but rather an information need that the visual presentation uniquely clarifies.

This should be a question that a user is asking this search system containing all of these Information Retrieval papers. It shouldn't be known in advance which particular paper contains this information.

For example, don’t ask “Which particular item was highlighted in bold?” or “Which particular item was colored red?”. These should questions and their answers should be about the information that is derived from the visual presentation, not the particular details of the visual presentation.

Remember, these should be questions you are asking a scientific literature mining system, you don't already know the particular paper.

If there are visuals like this, could you please write each such question and its ground truth answer that would only be possible to derive from the image? Could you also please mark which page these images are on that the question, answer pair is sourced from.

If there are not visuals like this, please just write "There are no visuals like this in the provided paper".
\end{quote}

\FloatBarrier

\section{Multimodal Hybrid Search with Closed-Source Models}
We extend our multimodal hybrid search analysis to closed-source embedding models. Figure 4 presents a comparison of fusion strategies using Voyage 3 Large for text embeddings and Cohere Embed v4.0 for image embeddings. Consistent with our open-source findings, Relative Score Fusion (RSF) generally outperforms Reciprocal Rank Fusion (RRF), though the optimal $\alpha$ value varies across recall depths.
\begin{figure}
    \centering
    \includegraphics[width=1.0\linewidth]{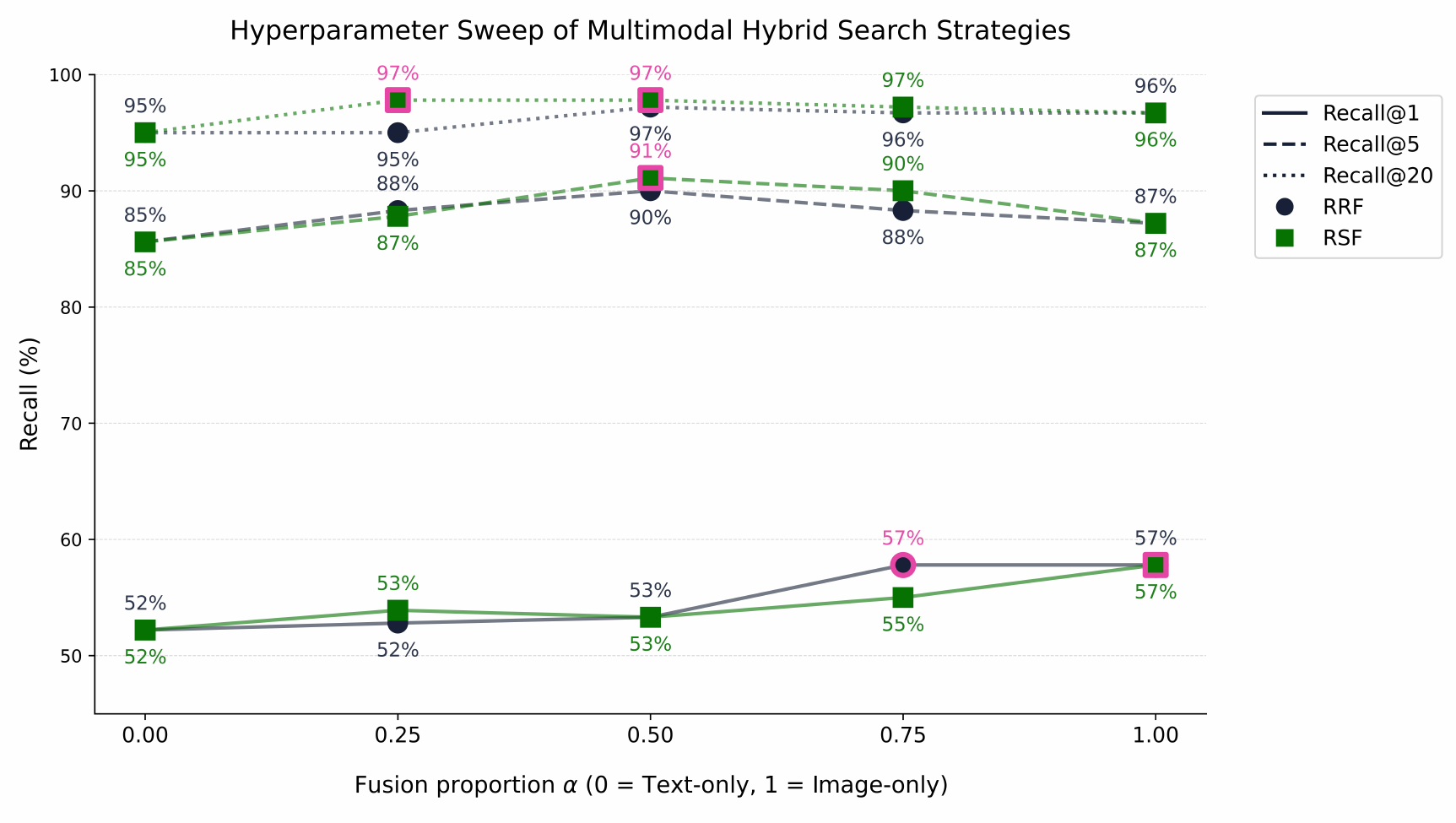}
    \caption{Comparison of Multimodal Hybrid Search fusion strategies with Hybrid Search using Voyage 3 Large and BM25 for Text and Cohere Embed v4.0 for Images. We compare Reciprocal Rank Fusion (RRF) and Relative Score Fusion (RSF) across different values of $\alpha$, where $\alpha$=0 uses text only and $\alpha$=1 uses images only. The highest performing configurations for each respective recall target (@1, @5, and @20) are highlighted in pink.}
    \label{fig:placeholder}
\end{figure}

\FloatBarrier

\section{Visualization of MaxSim Scoring}
To illustrate how late-interaction retrieval identifies relevant documents, we visualize the MaxSim scoring mechanism. Figure 5 presents a heatmap showing token-level similarity scores between a query and a retrieved page from the \textcolor{darkgreen}{IRPAPERS} corpus. High-attention regions correspond to query terms matching specific passages in the document, demonstrating how multi-vector representations enable fine-grained alignment between queries and document content.

\begin{figure*}
    \centering
    \includegraphics[width=0.8\linewidth]{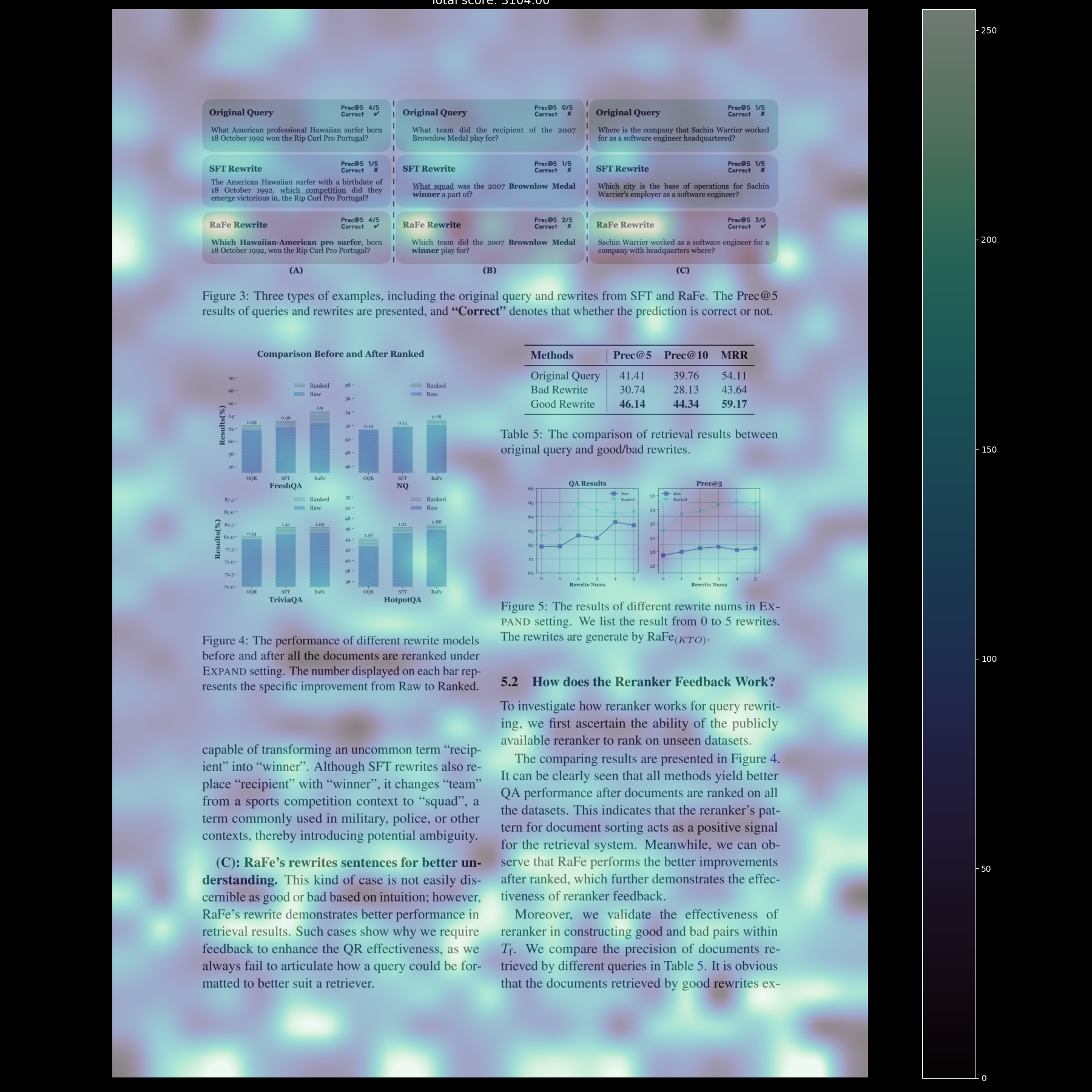}
    \caption{A heatmap visualization of MaxSim scoring for the query: \textit{What three categories does RaFe identify for how query rewriting improvements occur based on case studies?}}
    \label{fig:placeholder}
\end{figure*}

\FloatBarrier

\section{Additional Question, Answer Samples}
Table 6 provides additional examples of question-answer pairs from the \textcolor{darkgreen}{IRPAPERS} benchmark, illustrating the range of methodological details targeted by our evaluation. These samples demonstrate the specificity required to discriminate among semantically similar papers within the information retrieval literature.

\begin{table}[h]
\centering
\label{tab:csqe_qa}
\setlength{\tabcolsep}{12pt}
\renewcommand{\arraystretch}{1.5}
\begin{tabular}{|p{6cm}|p{8cm}|}
\hline
\textbf{Question} & \textbf{Answer} \\
\hline
What is the main limitation of LLM-generated query expansions that Corpus-Steered Query Expansion (CSQE) addresses in information retrieval? &
CSQE addresses misalignments between LLM-generated expansions and the retrieval corpus, including hallucinations, outdated information, and deficiency in long-tail knowledge due to limited intrinsic knowledge of LLMs. \\
\hline
In CSQE, how many LLM generations are sampled for KEQE expansions and corpus-originated expansions respectively? &
CSQE samples N=2 generations for both KEQE and corpus-originated expansions, totaling 4 generations, compared to N=5 for KEQE alone. \\
\hline
On the NovelEval dataset where LLMs lack knowledge, how does BM25+CSQE's nDCG@10 performance compare to BM25+KEQE? &
BM25+CSQE achieves 82.6 nDCG@10 compared to BM25+KEQE's 62.0, demonstrating CSQE's effectiveness when LLMs have no knowledge of the query topics. \\
\hline
Which LLM size shows CSQE can outperform supervised MS-MARCO-tuned dense retrieval models on TREC DL19? &
Using Llama2-Chat-70B, BM25+CSQE achieves 43.6 mAP, outperforming DPR (36.5), ANCE (37.1), and matching ContrieverFT (41.7) on DL19. \\
\hline
What prompt construction dataset does CSQE use for its one-shot learning context, and how does this affect its zero-shot classification? &
CSQE uses TREC DL19 dataset to construct the learning context in its prompt, making it zero-shot for all datasets except DL19 with minimal relevance supervision. \\
\hline
What is the MAP improvement range achieved by Generative Relevance Feedback (GRF) compared to RM3 expansion across different benchmark datasets? &
GRF improves MAP between 5-19\% compared to RM3 expansion across the evaluated benchmarks. \\
\hline
What are the 10 different generation subtasks that GRF uses to create query-specific text from large language models? &
The 10 subtasks are: Keywords, Entities, CoT-Keywords, CoT-Entities, Queries, Summary, Facts, Document, Essay, and News, with varying token lengths from 64 to 512 tokens. \\
\hline
How much does GRF improve NDCG@10 on the hardest 20\% of Robust04 topics compared to RM3's improvement over the baseline? &
GRF improves NDCG@10 by +0.145 on the hardest 20\% of topics, while RM3 only improves it by +0.006, representing a relative improvement of approximately 100-200\%. \\
\hline
\end{tabular}
\vspace{1em}
\caption{Question, Answer samples from \textcolor{darkgreen}{IRPAPERS}.}
\end{table}

\FloatBarrier

\section{t-SNE Visualization}
\label{app:visual_elements}

The t-SNE visualization found in HyDE \cite{hyde} is a perfect example of a visual that is difficult to convey with a text representation. The questions shown in Table 7 are designed to require this image representation in the reader model and highlight the respective limitations of text representations.

\begin{figure}[h]
\centering
\includegraphics[width=0.5\textwidth]{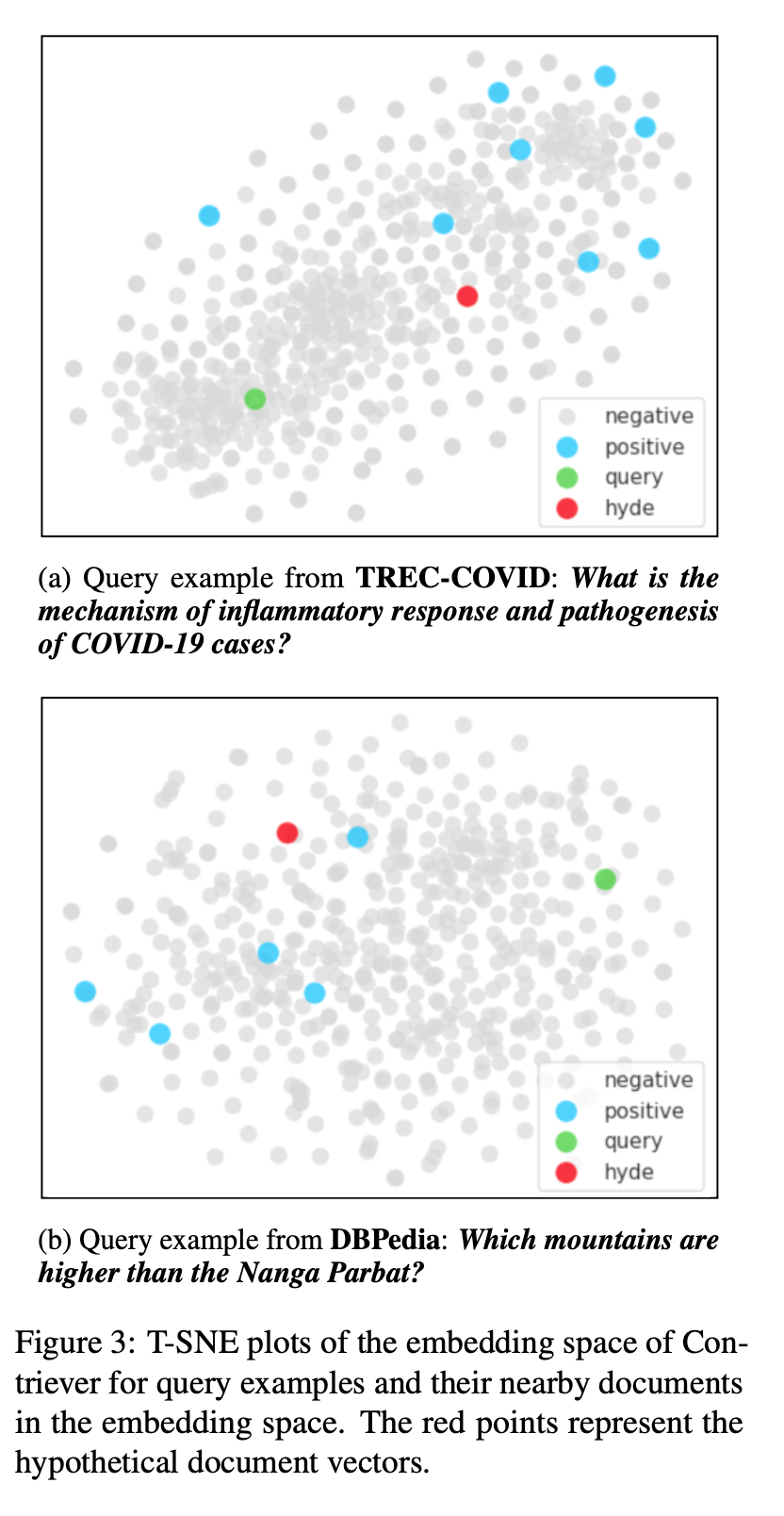}
\caption{Example abstract visualization showing t-SNE embeddings from HyDE \cite{hyde}. Unlike other visual categories, spatial relationships between points constitute the primary information content. Questions about cluster counts, relative positions, and geometric relationships cannot be reliably answered from text transcription, making this category uniquely suited to image-based representation.}
\label{fig:tsne_example}
\end{figure}

\FloatBarrier

\begin{table}[h]
\centering
\label{tab:tsne_questions}
\setlength{\tabcolsep}{8pt}
\renewcommand{\arraystretch}{1.5}
\begin{tabular}{|p{6.5cm}|p{7.5cm}|}
\hline
\textbf{Question} & \textbf{Answer} \\
\hline
How many distinct clusters of positive documents (cyan points) appear in the DBPedia visualization (bottom plot)? & 
The positive documents appear scattered across at least 3-4 distinct locations in the embedding space rather than forming a single tight cluster—there are cyan points on the far left, upper middle, right side, and lower regions. \\
\hline
In which of the two visualizations does the original query vector (green point) appear more isolated from all positive documents? & 
The TREC-COVID example (top plot) shows the query vector more isolated from positive documents, as the green point sits in the lower-center area while most cyan points cluster in the upper-right region. \\
\hline
Looking at the DBPedia example, is the HyDE vector closer to the nearest positive document or the original query vector? & 
The HyDE vector (red point) appears closer to at least one positive document (cyan point) nearby in the upper-middle region than to the original query vector (green point) positioned on the right side of the plot. \\
\hline
In the DBPedia example, are there any positive documents (cyan points) that appear to be outliers, far from both the query and HyDE vectors? & 
Yes, there is at least one positive document (cyan point) on the far left side of the DBPedia plot that appears isolated from both the query vector (green) on the right and the HyDE vector (red) in the upper-middle region. \\
\hline
\end{tabular}
\vspace{1em}
\caption{Sample questions targeting abstract visualizations (t-SNE plots) that require image-based representation. These questions cannot be reliably answered from text transcription alone.}
\end{table}

\end{document}